\documentclass[prl,twocolumn,amsmath,amssymb]{revtex4}

\newcommand{\nn}{\nonumber\\}
\newcommand{\ii}{{\rm i}}

\begin{document}

\title{Quantum mechanical retrodiction through an extended mean King problem}

\author{Amir Kalev}
\affiliation{Center for Quantum Information and Control, MSC07--4220, University of New Mexico,
Albuquerque, New Mexico 87131-0001, USA}

\author{Ady Mann and Michael Revzen}
\affiliation {Department of Physics, Technion - Israel Institute of Technology, Haifa
32000, Israel}

\date{\today}

\begin{abstract}
The mean King problem is a conditional retrodiction problem. In this problem Alice prepares
a two prime-dimensional particles state and avails one of the particles to the King who
measures its state in one of mutually unbiased bases of his choice. The King {\it tells}
Alice his choice of basis {\it after} she completes a control measurement on his particle.
Conditioned on this knowledge, she now infers the state observed by the King by utilizing
the outcome of her control measurement. In the {\it extended} mean King problem, studied in
this paper,  the King does not tell Alice his measurement basis, but instead both the King
and Alice repeat their measurements. Proper ordering of these allows
Alice to deduce {\it both} the basis used by the King and the outcome of his {\it first}
measurement, with the King reticent throughout, i.e., this
protocol effects a  complete (almost) retrodiction of the King's first
measurement.
\end{abstract}


\maketitle

{\it  Introduction}--- Retrodiction, viz. assigning the state of a system in the past, is
direct in deterministic classical physics: Knowledge of the present state, e.g. via
measurements, allows both prediction and retrodiction. Within quantum theory, the status of
this issue is not as clear as the theory is fundamentally a probabilistic theory.  Thus
e.g. does the theory retrodict  the observations or their probabilistic attributes? The
fundamental difficulty involved was noted almost at the inception of quantum mechanics
\cite{etp}: ``The principles of quantum mechanics actually involve an uncertainty in the
description of past events which is analogous to the uncertainty in the prediction of
future events.''  More recent studies pin the difficulty in quantum retrodiction on the
contextuality of quantum observables, suggested by Bell-Kochen-Specker theorem
\cite{mermin}.

The mean King problem (MKP), for which the present work is an extension, is a quantum
mechanical retrodiction problem \cite{werner,mermin}. It originated with \cite{lev1} for
spin-half particles and generalized to prime dimensionality in \cite{bg1} and power of
prime in \cite{arvind,durt,berge2}. Further generalizations were given in \cite{werner}. The MKP
involves a two-particle state prepared by Alice, availing one of the constituent particles
to the King who measures the particle's state, $m$, in one of mutually
unbiased bases of his choice, $b$ (unknown to Alice). Subsequent control measurement by Alice of
the two particles allows her to determine  -- {\it upon being informed} by the King of the
basis, $b$, he used -- the outcome, $m$, of his measurement, i.e., the state retrodiction is
`conditional' on further information: the King's measurement basis, $b$. Thus Alice
is {\it seemingly} \cite{mermin} violating a fundamental quantum tenant by being able to
assign (to one system) outcome values for non-commuting (distinct bases) observables.

In \cite{revz1,kmr} we considered a deterioration in the King's mood: Alice is {\it not}
being told the basis, $b$, (nor the outcome, $m$) of the King's measurement; we showed that
by an appropriate state preparation and measurements, she nonetheless is able to unravel
the basis, $b$. In this paper, we consider an extension of the MKP which, essentially,
completes the retrodiction. In the extended problem the King does not reveal his
measurement basis to Alice, but repeats his measurement (i.e. uses the same basis of the
first measurement). Alice, likewise, repeats her (control) measurement. Based on her
measurement results alone, Alice now retrodicts the King's {\it first} measurement: both
outcome and basis.  The protocol may be used as a key distribution with an extra
authentication scheme.

Mutually unbiased bases (MUB) play a central role in the analysis \cite{bg1,revz1,kmr}. In
what follows we list some  MUB characteristics that are utilized in the present work. Two
sets of orthonormal maximally entangled states studied in \cite{revz1} provide the
means for solving the retrodiction problem. The full protocol involved in retrodicting the
King's measurement is then spelled out. Furthermore we outline a possible
variant of the retrodiction protocol. The last part contains our conclusions inclusive of
possible implication of the results in quantum measurement theory.

{\it Brief Review of mutually unbiased bases}--- In a $d$-dimensional Hilbert space, two
complete, orthonormal bases, ${\cal B}_1,\;{\cal B}_2$,  are said to be MUB if and only if
(${\cal B}_1\ne {\cal B}_2)$
\begin{equation}
\forall |u\rangle\in{\cal B}_1\; {\rm and}\;\forall |v \rangle\in{\cal B}_2,\;\;|\langle
u|v\rangle|=1/\sqrt{d}.
\end{equation}
The physical meaning of this is that knowledge that a system is in a particular state in
one basis implies complete ignorance of its state in the other basis.

Ivanovic \cite{ivanovic} proved that there are at most $d+1$ MUB  in a
$d$-dimensional Hilbert space and gave an explicit formula for the $d+1$ bases in the case
of $d$=prime number. Wootters and Fields \cite{wootters2} constructed such $d+1$ bases for
power of prime dimensions. Variety of methods for construction of the $d+1$ bases for
power of prime dimensions are now known \cite{tal,wootters3,klimov2,vourdas}.

Our present study is confined to $d$=prime number. We now give explicitly the
MUB states in conjunction with the Weyl-Schwinger pair $Z,X$
\cite{schwinger,amir,rev1}. Thus we label the $d$ orthonormal states spanning the Hilbert
space,  termed  the computational basis, by $|n\rangle$, $n=0,\ldots,d-1$, and
$|n+d\rangle=|n\rangle$,
\begin{align}
Z|n\rangle&=\omega^{n}|n\rangle,\;\omega=e^{\ii 2\pi/d},\nn
X|n\rangle&=|n+1\rangle.
\end{align}
The above relation, together with the commutation relation
\begin{align}\label{commrel}
ZX=\omega XZ,
\end{align}
imply that the pair $Z,X$ completely specify the degree of freedom. The $d$ states in each
of the $d+1$ MUB bases \cite{tal,amir} are the computational basis  and the $d$ bases:
\begin{equation} \label{mxel}
|m;b\rangle=\frac{1}{\sqrt d}\sum_0^{d-1}\omega^{\frac{b}{2}n(n-1)-
 nm}|n\rangle,\;\;b,m=0,\ldots,d-1,
\end{equation}
where $b$ labels the bases, and $m$ labels the states within a basis. Each basis
corresponds to Weyl-Schwinger pair $Z,X$ by \cite{tal}
\begin{equation}\label{tal1}
XZ^b|m;b\rangle=\omega^m|m;b\rangle.
\end{equation}
For later reference we shall refer to the computational basis by $b=\ddot{0}$. We denote
$|m;\ddot{0}\rangle$ by $|m\rangle$ when no confusion should arise. Thus
the $d+1$ bases are labeled as $b=\ddot{0},0,1,\ldots,d-1$. We have of course,
\begin{equation}\label{mub}
|\langle m;b|m';b'\rangle|=\delta_{m,m'}\delta_{b,b'}+(1-\delta_{b,b'})\frac{1}{\sqrt d}.
\end{equation}
This completes our discussion of single particle MUB.

Several studies  \cite{fivel1,durt1,klimov1,berge2,ent} considered the entanglement of two $d$-dimensional
particles via MUB state labeling. Our presentation is based on
\cite{revz1,kmr}. In \cite{revz1,kmr} we considered maximally entangled states that are {\it product} states in
the particles' collective (i.e., ``center of mass'' and ``relative'') coordinates. Based on
these latter representative states one obtains
\cite{revz1} orthonormal bases that span the $d^2$ dimensional Hilbert space with maximally entangled states. In
the next section we use two such bases sets.

{\it Retrodiction of the King's measurement}--- Retrodiction within quantum mechanics
involves the specification of past measurement outcome. To effect such retrodiction the
extended MKP includes the following bases,
\begin{align}\label{meas}
&\{|m;b\rangle|m=0,\ldots,d-1\}\;\;{\rm single\;particle\;MUB},\nonumber \\
&\{|u,v;-\rangle|u,v=0,\ldots,d-1\}\;\;{\rm two{-}particle\;basis},\nonumber \\
&\{|u,v;+\rangle|u,v=0,\ldots,d-1\}\;\;{\rm two{-}particle\;basis}.
\end{align}
The last two bases (the `minus' and `plus' bases) are manifestly orthonormal sets of maximally entangled states
for the two $d$-dimensional particles \cite{revz1,kmr}:
\begin{align}\label{states}
|u,v;-\rangle&=\frac{1}{\sqrt
d}\sum_{n=0}^{d-1}|n\rangle_1X_2^{2u}Z_2^{v}|-n\rangle_2 \nonumber \\
|u,v;+\rangle&=\frac{1}{\sqrt d}\sum_{n=0}^{d-1}|n\rangle_1X_2^{-2u}Z_2^{-v} |n\rangle_2.
\end{align}

A possible protocol is as follows. Alice prepares a state from the `plus' basis, say
$|u,v;+\rangle$, cf. Eq.(\ref{states}), and sends particle 1 to the King. As promised, the
King measures the particle in one of the $d+1$ MUB of his choice. His outcome is, say, $m$,
and he returns the particle to Alice. Now Alice measures the two
particles in the `minus' basis, $\{|c,r;-\rangle\;|c,r=0,\ldots,d-1\}$ and observes, say,
an outcome labelled by $u'$ and $v'$. Thus, by direct evaluation of
\begin{equation}
_{1,2}\langle u',v';-|m;b\rangle_1\langle m;b|u,v;+\rangle_{1,2} \ne 0,
\end{equation}
she obtains,
\begin{align}\label{m(b)}
m=\begin{cases}
    u'+u & \text{for } b= \ddot{0}\\
   \frac{v'+v}{2}+b(u'+u) -\frac{b}{2}& \text{for } b= 0,\ldots, d-1.
  \end{cases}
\end{align}
Had Alice been informed of the basis used by the King, $b$, she could have deduced his
outcome, $m$. However in this protocol the King does not disclose his measurement basis.
Instead Alice now sends particle 1 again to the King who repeats his measurement (i.e.,
uses the same measurement basis $b$), obtains some result, say $m'$ and returns the
particle to Alice. Actually,  the outcome of his second measurement is immaterial for the
protocol \cite{lajos,kmr}. The King might as well use a nonselective measurement in the $b$
basis \cite{kmr}. At this point, Alice repeats her measurement in the `minus' basis, and
obtains some result, $u'',v''$. This allows her to deduce the basis used (twice) by the
King. Thus, the condition
\begin{equation}
_{1,2}\langle u'',v'';-|m';b\rangle_{1}\langle m';b|u',v';-\rangle_{1,2} \ne 0,
\end{equation}
implies that
\begin{align}\label{findb}
b=\begin{cases}
    \ddot{0} & \text{for } u''=u'\\
   \frac{v'-v''}{2(u'-u'')} & \text{for } \;u''\ne u'.
  \end{cases}
\end{align}
We note that the above procedure fails when {\it both} $u''=u'$ {\it and} $v''=v'$, in which
case the basis $b$ is undetermined and Alice is unable to retrodict the King's observation.
This case happens with probability $1/d$. Upon learning the measurement basis, Alice uses
Eq.~(\ref{m(b)}) to find the outcome of the King's first measurement as well. This
completes the retrodiction.

An obvious alternative protocol would be a retrodiction of the King's {\it second} measurement (assumed {\it selective} in this case). This requires a simpler procedure: Alice prepares her state,
e.g. $|u,v;+\rangle$, the King measures (non selectively)  in one of the MUB. Now Alice measures
in the `plus' basis. This allows her to deduce \cite{kmr} the basis, $b$, used by the King. The
King's repeated measurement leaves particle 1 in the state $|m,b\rangle$, thus allowing
Alice to determine $m$ simply by measuring this particle in the same basis used by the King.

The above protocol was formulated for an odd prime dimension. The mathematical reason for
that may be traced back to Eqs.~(\ref{mub}) and (\ref{states}). We can,
however, include the dimension two as well. In this case, the first part of the protocol
follows the original protocol of \cite{lev1}, by which Alice obtains the information
$m(b)$, that is the outcome of the King's measurement conditioned on his measurement basis.
Here, as before, the King does not reveal his measurement basis, but instead Alice prepares
another state, say $|u',v';+\rangle$ in the `plus' basis, formed by the states
\begin{equation}\label{d2 states}
|u,v;+\rangle=\frac{1}{\sqrt
2}\sum_{n=0,1}|n\rangle_1X_2^{u}Z_2^{v}|n\rangle_2,\; u,v=0,1,
\end{equation}
and sends particle 1 to the King. He, in turn, measures the particle in the same basis he
used in his first measurement and returns the particle to Alice, who measures the two
particles in the `plus' basis of Eq.~(\ref{d2 states}). From her measurement results
$u'',v''$ she can deduce the measurement basis used by the King according to
\begin{align}\label{d2 findb}
b=\begin{cases}
    \ddot{0} & \text{for } u''=u'\\
   \frac{v'-v''}{u'-u''} & \text{for } \;u''\ne u'.
  \end{cases}
\end{align}
The procedure fails when {\it both} $u'=u''$ {\it and} $v'=v''$, in which case the basis
$b$ is undermined and Alice is unable to retrodict the King's observation. This happens
with probability 1/2. (Hence, considering $n$ two-level systems, the
probability for the undetermined case would be $1/2^n$.)

{\it Conclusions and Discussion}--- Classical physics being deterministic and time
symmetric allows direct retrodiction: measuring an observable implies its value at an
earlier time. Quantum mechanics, though time symmetric, having a distinct approach to
measurements and being a probabilistic theory cannot directly be classified as allowing
retrodiction. Thus, e.g., Bell-Kochen-Specker theorem seems to require contextuality among
the observables, straining retrodiction.  We demonstrated above that, using entanglement
and extension of a protocol used for tracking measurement bases, it is possible to
retrodict, i.e.,  to give the basis and outcome of past measurement given present
experimental values under certain specified conditions.

The study may be summarized as follows:  The King measured some observables associated with
a measurement setting, $b$ (in our case $b=\ddot{0},0,1,\ldots,d-1$) and obtained an
outcome $m$ (in our case $m=0,\ldots,d-1$); Alice via her control measurement can tabulate
the correspondence $m(b)$ for all $b$. Thus, within classical physics, being informed of
$b$ she can stipulate (for both past and future) the King's outcome, $m$. We may refer to
this as `conditional' retrodiction, it being conditioned on knowing the ``alignment'', $b$.
On the other hand, quantum mechanics precludes such definite outcome values in general and
particularly when distinct alignments correspond to non commuting observables (as is in our
case the MUB $b$ correspond to $XZ^b$, $[XZ^{b'},XZ^b]\ne0,\;b \ne b'$). Nonetheless, the
corresponding tabulation is provided within the solution to the MKP. This is possible only
as `conditional' retrodiction with the {\it understanding} that the tabulated value,
$m(b)$, has a meaning only for the actually chosen alignment $b$. The other are
meaningless (this point was stressed in \cite{mermin}).

Now in the extended MKP studied in this work the King repeats his measurement after Alice
undertook her (first) control measurement. Within classical physics this does {\it not}
provide any extra information to Alice, in particular as the King is allowed to ignore the
outcome of his second measurement (i.e., he can perform a nonselective measurement).
However, within quantum mechanics this may, as is shown in this work, provide enough extra
information to allow (almost) complete retrodiction of his (first) measurement, that is, it
allows Alice to deduce {\it both} the ``alignment", $b$, used by the King and the outcome,
$m$, of his measurement.

The study may be viewed as providing a novel means for a key distribution within a
cryptographic protocol. Thus the parties may agree on having both/either the outcome of the King
(first) measurement and/or the basis he used as forming their key and use the other
observable for checking on the security of their communication.

AK would like to acknowledge the support of NSF Grants No. PHY-1212445.

\end{document}